\documentclass[a4paper,twoside,notoc,11pt]{JHEP3}

\usepackage{epsfig,multicol}
\usepackage{delarray,amsmath,bbm}

\newcommand\fverb{\setbox\pippobox=\hbox\bgroup\verb}
\newcommand\fverbdo{\egroup\medskip\noindent%
                              \fbox{\unhbox\pippobox}\ }
\newcommand\fverbit{\egroup\item[\fbox{\unhbox\pippobox}]}
\newbox\pippobox

\newcommand{\nn}{\nonumber}
\newcommand{\beq} {\begin{equation}}
\newcommand{\eeq} {\end{equation}}
\newcommand{\beqa} {\begin{eqnarray}}
\newcommand{\eeqa} {\end{eqnarray}}

\newcommand{\ie}{{\it i.e.}}
\newcommand{\eg}{{\it e.g.}}
\newcommand{\cf}{{\it cf.}}
\newcommand{\etal}{{\it et al.}}

\newcommand{\ieps}{i\varepsilon}

\newcommand{\morder}[1]{{\cal O}\left(#1 \right)}
\newcommand{\eq}[1]{(\ref{#1})}

\newcommand{\inv}[1]{\frac{1}{#1}}

\newcommand{\qu}{{\rm q}}

\newcommand{\halft}{{\textstyle \frac{1}{2}}}

\newcommand{\aslash}[1]{ \rlap{/}{#1} }

\newcommand{\Slash}[1]{ \parbox[b]{0.6em}{$#1$} \hspace{-0.55em}
                       \parbox[b]{0.55em}{\raisebox{-0.2ex}{$/$}}}
                       
\newcommand{\bs}{\boldsymbol}
\newcommand{\mM}{\mathcal{M}}
\newcommand{\mN}{\mathcal{N}}
\newcommand{\mA}{\mathcal{A}}
\newcommand{\mB}{\mathcal{B}}
\newcommand{\im}{{\rm Im}}
\newcommand{\re}{{\rm Re}}

\newcommand{\kbf}{\bs{k}}

\newcommand{\rbf}{\bs{r}}

\title{\center{Soft Rescattering in DIS: Effects of Helicity Flip}}

\author{Paul Hoyer$^{a,b}$ and Matti J\"arvinen$^a$\\
              $^a$Department of Physical Sciences and Helsinki Institute of
              Physics\\
              \ POB 64, FIN-00014 University of Helsinki, Finland \\
              $^b$NORDITA, Blegdamsvej 17, DK-2100 Copenhagen, Denmark\\
              E-mails: \email{paul.hoyer@helsinki.fi, Matti.O.Jarvinen@Helsinki.fi}}
\preprint{HIP-2005-36/TH \\NORDITA-2005-56 \\  \hepph{0509058} \\ \today
}

\abstract{Soft rescattering in hard QCD processes may involve non-perturbative, chiral symmetry breaking interactions. We find that rescattering which changes the helicity of the struck quark gives rise to a leading twist single spin asymmetry in semi-inclusive DIS, the angular dependence of which is the same as that usually ascribed to the Collins effect.
We argue that helicity-changing rescattering can contribute also to $k_\perp$-integrated parton distributions.
}

\keywords{Deep Inelastic Scattering, QCD, Spin and Polarization Effects}

\begin{document}

\section{Introduction}

Deep Inelastic Scattering (DIS, $e+N \to e+X$) is non-local on the light-cone ($x^2=0$). Some time ago it was noted \cite{Brodsky:2002ue,Brodsky:2004hi} that soft rescattering occurring within the finite Ioffe coherence length along the light-cone has physical, observable effects such as shadowing and diffraction. Studies of pertubative model amplitudes have led to
broad agreement that such rescattering enables a single spin asymmetry ($A_N$) at leading twist \cite{Brodsky:2002cx,Collins:2002kn}, the ``Sivers'' effect \cite{Sivers:1989cc} in semi-inclusive DIS. In the model proposed by Brodsky, Hwang and Schmidt (BHS, \cite{Brodsky:2002cx}) a transverse spin asymmetry arises from the interference of the Born term with the absorptive part of the one-loop amplitude, and is thus proportional to an elastic, on-shell rescattering amplitude of the struck quark. It is recognized that such calculations are not quantitatively reliable due to the softness of the rescattering. However, it appears unlikely that a non-vanishing effect thus established would be cancelled by the non-perturbative sector of QCD.

In this paper we consider what effects the specifically non-perturbative features of QCD may have on $A_N$. Due to spontaneous chiral symmetry breaking the helicity of massless quarks need not be conserved in soft processes. {\it E.g.,} instantons can generate a Pauli-type coupling (anomalous magnetic moment) of quarks to gluons \cite{Kochelev:1996pv}. We use the BHS model to study the effect on $A_N$ of adding a Pauli coupling at the vertex between the struck quark and the exchanged gluon,
\beq \label{paulivert}
 -ig_s\gamma ^\mu \to -ig_s\gamma^\mu + a(p^2)\sigma^{\mu\nu}p_\nu
\eeq
where the effective coupling $a(p^2)$ vanishes at large virtualities $p^2$ of the gluon.

We present our calculation in the next section. However, one can qualitatively anticipate the result. The remarkable fact that a soft gluon can be coherent with the hard photon (of virtuality $Q^2 \to \infty$) is a consequence of Lorentz dilation. The proper coherence length $1/Q$ of the virtual photon is dilated by a factor $\gamma = \nu/Q$ in the target rest frame, where the photon has energy $\nu$. Thus the Ioffe coherence length is finite in the Bjorken limit, and inversely proportional to $x_B$,
\beq \label{ioffe}
L_I = \inv{Q} \cdot \frac{\nu}{Q} = \inv{2Mx_B}
\eeq
where $M$ is the target mass. Note that soft interactions with partons that are {\it comoving} with the struck quark (\ie, which emerge in the current fragmentation region), do not benefit from the dilation factor and thus are incoherent with the hard process. This is why hadronization does not influence the DIS cross section at leading twist.

For the rescattering to be relevant its cross section must be constant in the high energy ($\nu \to \infty$) limit, \ie, the rescattering amplitude must be $\propto \nu$. This is indeed the case for longitudinal gluon exchange, whereas transverse gluon exchange does not contribute. It is readily checked that also the Pauli part of the vertex in \eq{paulivert} gives an elastic amplitude $\propto \nu$. This leads one to expect that the Pauli coupling will contribute to $A_N$, which is caused by an interference between amplitudes where the target has opposite helicity. As we shall see, the azimuthal dependence of this new contribution is the same as that of the Collins effect \cite{Collins:1992kk}.

\section{Single spin asymmetry with Pauli coupling in the BHS model}

\begin{figure}[hbt]
\begin{center}
\epsfig{figure=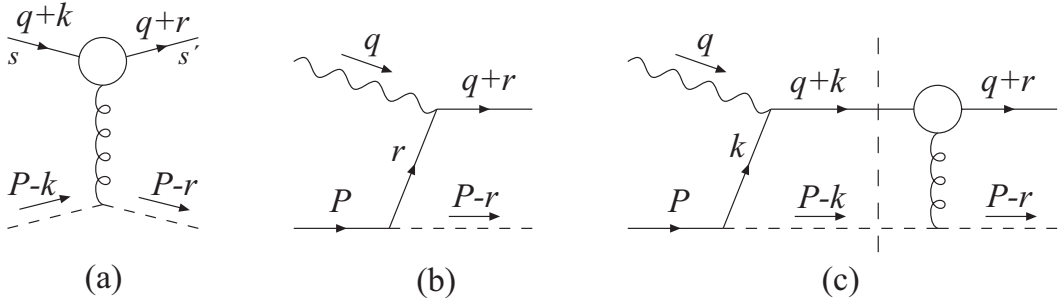,angle=0,width=14cm}
\end{center}
\caption{Amplitudes which contribute to the asymmetry. The blob represents the Pauli coupling in (1.1). (a) Elastic scattering amplitude with Pauli coupling. (b) 
The Born amplitude. (c) The discontinuity of the loop amplitude.
\label{ampfig}}
\end{figure}

We study the single transverse-spin asymmetry in semi-inclusive DIS using the BHS model \cite{Brodsky:2002cx}, in which a target quark emits a scalar particle before being struck by the virtual photon (\cf\ Fig. \ref{ampfig}). We
evaluate the effect of an anomalous magnetic moment (Pauli coupling) in the rescattering of the struck quark. Such a coupling might be generated non-perturbatively, due to the softness of the rescattering gluon.

The spin asymmetry for a target polarized in the transverse ($y$) direction is defined as
\beq \label{asymm}
 A_N \equiv \frac{\sum_{\{\sigma\}}\left[|\mM_{\uparrow,\{\sigma\}}|^2-|\mM_{\downarrow,\{\sigma\}}|^2 \right]}{\sum_{\{\sigma\}}\left[|\mM_{\uparrow,\{\sigma\}}|^2+|\mM_{\downarrow,\{\sigma\}}|^2  \right]} 
= \frac{2 \sum_{\{\sigma\}} \im\left[\mM_{\leftarrow,\{\sigma\}}^*\mM_{\rightarrow,\{\sigma\}}\right]}{\sum_{\{\sigma\}}\left[|\mM_{\rightarrow,\{\sigma\}}|^2+|\mM_{\leftarrow,\{\sigma\}}|^2  \right]}
\eeq 
where the $\mM_{\leftrightarrow,\{\sigma\}}$ are helicity amplitudes and $\{\sigma\}$ represents the helicities of all particles except the target.
The asymmetry vanishes at Born level but can get a contribution from the interference between the Born amplitude and the absorptive part of the one-loop amplitude. We study the asymmetry which arises when a Pauli coupling is used at the quark-gluon vertex of the rescattering in Fig. \ref{ampfig}(c).

We introduce the Pauli coupling through the replacement \eq{paulivert} at the quark-gluon vertex, where $p$ is the (incoming) gluon momentum. In the absorptive part of the rescattering amplitude the quark is on-shell. If the Pauli contribution arises from non-perturbative effects $a(p^2)$ should vanish at large virtualities $-p^2$ of the exchanged gluon. For our purposes the precise dependence on $p^2$ is not important, we take
\beq \label{paulipar}
 a(p^2) = a_0 e^{ A p^2}
\eeq
where $a_0$ and $A$ are constants.

We use a coordinate system where the virtual photon momentum is along the $+z$ axis and the target is at rest, \ie, the photon and target momenta read
\beqa \label{csyst}
 q &=& (q^+,q^-,0_\perp) \simeq  (2\nu,-x_B M,\bs{0}_\perp)
 \nn\\
 P &=& (M,M,\bs{0}_\perp) 
\eeqa
where $\nu$ is the photon energy, $M$ is the target mass and $x_B\equiv Q^2/2 M \nu$ is the Bjorken variable. The struck quark thus gets a large momentum in the $+z$ direction, making the use of the standard light-front (LF) spinors defined in \cite{Brodsky:1997de} convenient. We evaluate the asymmetry $A_N$ at leading twist with finite target and scalar masses $M$ and $m_s$, taking the quark mass $m$ to be zero.

\subsection{The amplitudes}

The relevant amplitudes are shown in Fig. \ref{ampfig}. The Pauli coupling contribution to the on-shell elastic quark-scalar amplitude (Fig. \ref{ampfig}(a)) is given by
\beqa
 i \mM^{\textrm{el}}_{ss'} &\equiv& \bar u(q+r,s') a_0 e^{A(k-r)^2}\sigma^{\mu\nu}u(q+k,s) (r_\nu-k_\nu)(2 P_\mu-k_\mu-r_\mu)\nn\\ 
	&\times& \frac{-i}{(k-r)^2}(-ie_2)
\eeqa
with the LF spinors defined in \cite{Brodsky:1997de}.
The helicity of the (massless) quark flips at the Pauli vertex: only the spin flip amplitudes are non-zero. Their leading behavior in the limit of $q^+ \to\infty$ at fixed $k,r$ is
\beq \label{elamp}
 \mM^{\textrm{el}}_{+-} \simeq 2 a_0 e_2 \frac{(1-x_B) M q^+\left(k_\perp e^{+i\phi}-r_\perp e^{+i\psi}\right)}{(\kbf_\perp-\rbf_\perp)^2}
e^{-A(\kbf_\perp-\rbf_\perp)^2}\simeq -\left(\mM^{\textrm{el}}_{-+}\right)^*
\eeq 
where we used $k^-\simeq r^- \simeq M x_B$ (as required for the quarks to be on-shell, with $q^- \simeq -Mx_B$) and denoted
\beqa \label{quang}
 \kbf_\perp &=& k_\perp (\cos \phi,\sin\phi)
\nn\\
 \rbf_\perp &=& r_\perp (\cos \psi,\sin\psi)
\eeqa
Note that the amplitudes (\ref{elamp}) are proportional to 
$s \simeq M q^+$.
This is required for the Pauli coupling to contribute at leading twist in 
Fig.~\ref{ampfig}(c).

The Born amplitude of Fig. \ref{ampfig}(b) reads
\beq
 i\mA_{ss'}^\lambda \equiv \bar u(q+r,s')(-ie_1 \Slash{\varepsilon}^\lambda)\frac{i\aslash{r}}{r^2}(-ig) u(P,s) 
\eeq
where $u(P,s)$ is the target spinor (of mass $M$).
At leading twist only transversely polarized photons contribute, with polarization vectors 
$\varepsilon^{\lambda=\pm 1} \equiv -(0,0,\pm 1,i)/\sqrt{2}\,$. We get\footnote{Here the spin non-flip amplitudes are proportional to the proton mass $M$ whereas in the
original BHS model they were $\propto r_\perp$. This is due to our choice of frame (\ref{csyst}).} 
\beqa \label{bornamp}
 \mA_{++}^\lambda &\simeq& e_1 g \sqrt{2M q^+} \, \frac{1-x_B}{r_\perp^2+B}\, r_\perp e^{+i\psi}\, \delta_{\lambda,+1} 
\nn\\
 \mA_{+-}^\lambda &\simeq& e_1 g \sqrt{2M q^+} \, \frac{1-x_B}{r_\perp^2+B}\,Mx_B 
 \, \delta_{\lambda,-1} 
\eeqa
where 
\beq \label{Bexpr}
B \equiv x_B \left[m_s^2-(1-x_B)M^2\right]
\eeq
The remaining helicity amplitudes are obtained through
\beq
 \mA_{ss'}^\lambda = -(-1)^{s-s'} \left(\mA_{-s,-s'}^{-\lambda}\right)^*
\eeq
where $s,s' = \pm \halft$. The amplitudes $\mM$ in \eq{asymm} involve also the contribution from the lepton vertex. The denominator of \eq{asymm} is (at lowest order) given by the Born amplitude through
\beq \label{norm}
\mN \equiv \sum_{\{\sigma\}}\left[|\mM_{\rightarrow,\{\sigma\}}|^2+|\mM_{\leftarrow,\{\sigma\}}|^2  \right] = \inv{Q^4} \sum_{s,s'}\sum_{\lambda,\lambda'} L^{\lambda,\lambda'} \mA_{ss'}^\lambda {(\mA_{ss'}^{\lambda'})}^*
\eeq
Defining the transverse momentum of the leptons as
\beq \label{lepang}
\bs{l}_{1\perp} = \bs{l}_{2\perp} = l_\perp(\cos\tau,\sin\tau)
\eeq
the dependence on the lepton momenta $l_1,l_2$ is given by
\beq \label{Lexpr}
L^{\lambda,\lambda'} = \frac{2e^2 Q^2}{y^2} \left\{\left[1+(1-y)^2\right]\delta_{\lambda,\lambda'} - 2(1-y) e^{-2i\lambda\tau}\delta_{\lambda,-\lambda'} \right\}\ \ \mbox{for } \lambda, \lambda'=\pm 1
\eeq
where $y \equiv P \cdot q/P \cdot l_1$ is the fraction of the beam energy carried by the virtual photon. The helicity dependence of the Born amplitudes \eq{bornamp} restricts the photon helicities in \eq{norm} to $\lambda = \lambda' =\pm 1$. Using \eq{Lexpr} we find
\beq \label{normexpr}
\mN = \frac{8e^2}{y^2 Q^2} \left[1+(1-y)^2\right] (e_1g)^2 Mq^+ \left(\frac{1-x_B}{r_\perp^2+B}\right)^2 \left[ r_\perp^2 + (Mx_B)^2\right]
\eeq

The loop amplitude is shown in Fig. \ref{ampfig}(c). Only its discontinuity (absorptive part) contributes to the asymmetry \eq{asymm}. According to the Cutkosky rules the discontinuity is given by a convolution of the amplitudes (\ref{elamp}) and (\ref{bornamp}):
\beqa \label{loopamp}
 \textrm{Disc}\, \mB_{ss'}^\lambda &\equiv& i \int \frac{d^4 k}{(2\pi)^4}\ 2 \pi \delta\left[(k+q)^2\right]\ 2 \pi \delta\left[(P-k)^2-m_s^2\right] 
 \mA_{s,-s'}^\lambda \mM_{-s',s'}^\textrm{el} \nn\\
  &\simeq& \frac{i}{2(1-x_B)M q^+} \int \frac{d^2 \kbf_\perp}{(2\pi)^2} \mA_{s,-s'}^\lambda \mM^{\textrm{el}}_{-s',s'} = (-1)^{s-s'} \left(\textrm{Disc}\, \mB_{-s,-s'}^{-\lambda} \right)^*
\eeqa
where $\mA_{s,-s'}^\lambda$ is to be expressed in terms of the quark momentum $k$ rather than $r$ as in \eq{bornamp}.

\subsection{Evaluation of the asymmetry}

The single spin asymmetry (\ref{asymm}) is given by the amplitudes (\ref{bornamp}) and (\ref{loopamp}) as
\beqa \label{paulias}
 \mN A_N  &\simeq& \frac{2}{Q^4} \sum_{\lambda,\lambda'} \sum_{s'} \im \left\{ L^{\lambda,\lambda'}\left[\mA_{+,s'}^{\lambda}+\halft \textrm{Disc}\, \mB_{+,s'}^{\lambda}\right]
 \left[\mA_{-,s'}^{\lambda'}+ \halft\textrm{Disc}\, \mB_{-,s'}^{\lambda'}\right]^* \right\}
\nn\\
&=& \frac{8e^2}{Q^2} \frac{1-y}{y^2} \sum_{\lambda=\pm 1} \im \left\{ \lambda e^{-2i\lambda\tau} \mA_{+,\lambda/2}^{\lambda}\, \textrm{Disc}\, \mB_{+,-\lambda/2}^{\lambda}\right\}
\eeqa
where the contribution arises from photons with opposite helicities, $\lambda = - \lambda'$. From the explicit expressions of the amplitudes we get
\beq
\mN A_N = \frac{16 e^2}{Q^2} \frac{1-y}{y^2} (e_1g)^2 e_2 a_0 Mq^+ \frac{(1-x_B)^2}{r_\perp^2+B} \re \left\{ e^{-2i\tau}I^+ + e^{2i\tau} I^- \right\}
\eeq
where 
\beqa \label{Iplus}
 I^+  &\equiv&\int \frac{d^2 \kbf_\perp}{(2\pi)^2}
\frac{k_\perp e^{i\phi}\left(k_\perp e^{i\phi} -r_\perp e^{i\psi}\right)r_\perp e^{i\psi}}{(k_\perp^2+B)(\kbf_\perp-\rbf_\perp)^2}e^{-A(\kbf_\perp-\rbf_\perp)^2}\nn\\
&=& e^{3i\psi}\int \frac{d^2 \kbf_\perp}{(2\pi)^2}
\frac{\left(k_\perp e^{i\phi}+r_\perp \right)k_\perp e^{i\phi} r_\perp\ e^{-Ak_\perp^2} }{\left(k_\perp^2+r_\perp^2+2 k_\perp r_\perp \cos\phi +B\right)k_\perp^2}
 \equiv e^{3i\psi} J^+
\eeqa
and we substituted $\kbf_\perp \to \kbf_\perp+\rbf_\perp$ followed by $\phi \to \phi + \psi$ in \eq{Iplus}. Similarly,
\beqa \label{Iminus}
I^- &\equiv& \int \frac{d^2 \kbf_\perp}{(2\pi)^2}
\frac{(Mx_B)^2\left(k_\perp e^{-i\phi} -r_\perp e^{-i\psi}\right)}{(k_\perp^2+B)(\kbf_\perp-\rbf_\perp)^2}e^{-A(\kbf_\perp-\rbf_\perp)^2} \nn\\
 &=& e^{-i\psi} \int \frac{d^2 \kbf_\perp}{(2\pi)^2}
\frac{(Mx_B)^2 k_\perp e^{-i\phi} \ e^{-Ak_\perp^2} }{\left(k_\perp^2+r_\perp^2+2 k_\perp r_\perp \cos\phi +B\right)k_\perp^2}
\equiv e^{-i\psi} J^- 
\eeqa
with $B$ given by \eq{Bexpr}.
Noting that the $J^\pm$ are real and using the expression \eq{normexpr} for $\mN$ gives for the asymmetry
\beq
 A_N \simeq 2 e_2 a_0 \frac{1-y}{1+(1-y)^2} \frac{r_\perp^2+x_B \left[m_s^2-(1-x_B)M^2\right]}{\left[r_\perp^2+(Mx_B)^2\right]} \left[J^+ \cos(3\psi-2\tau)+ J^- \cos(\psi-2\tau) \right]
\eeq

The integrals in the factors $J^\pm$ defined in \eq{Iplus} and \eq{Iminus} can be evaluated explicitly in the limit where the rescattering is very soft, \ie, for $r_\perp^2,m_s^2,M^2 \gg 1/A$, where $A$ determines the momentum dependence of the Pauli coupling \eq{paulipar}. The result for the asymmetry then reads
\beqa \label{result}
 A_N &\simeq& - \frac{e_2 a_0}{2 \pi A} \frac{1-y}{1+(1-y)^2} \frac{r_\perp}{\left\{r_\perp^2+x_B \left[m_s^2-(1-x_B)M^2\right]\right\} \left[r_\perp^2+(Mx_B)^2\right]}
\nn\\
\\
&\times& \left[r_\perp^2\cos(3\psi-2\tau)+ (M x_B)^2 \cos(\psi-2\tau)\right]
 \left[1+\morder{\frac{1}{A}} \right] \nn
\eeqa
Note that $A_N \propto 1/r_\perp$ for large $r_\perp$.

\section{Concluding remarks}

In the preceding section we used a frame where the target spin was fixed in the $y$ direction, with quark and lepton azimuthal angles defined by Eqs. (\ref{quang}) and (\ref{lepang}), respectively. In the Trento convention \cite{Bacchetta:2004jz} one uses instead the angle 
$\phi_s$ between the target spin direction and the lepton plane,
\beq
 \phi_s = \pi/2-\tau
\eeq
and the angle $\phi_h$ between the hadron (or quark jet) and lepton planes
\beq 
 \phi_h = \psi-\tau 
\eeq
In terms of these angles our result (\ref{result}) is
\beqa \label{trres}
 A_N &\simeq&
\frac{e_2 a_0}{2 \pi A} \frac{1-y}{1+(1-y)^2} \frac{r_\perp}{\left\{r_\perp^2+x_B \left[m_s^2-(1-x_B)M^2\right]\right\} \left[r_\perp^2+(Mx_B)^2\right]}
\nn\\
\\
&\times& \left[r_\perp^2\sin(3\phi_h-\phi_s)-(Mx_B)^2 \sin(\phi_h+\phi_s)\right]
\left[1+\morder{\frac{1}{A}} \right] \nn
\eeqa

The angular dependence is the same\footnote{Except that our angle $\phi_h$ refers to the current jet axis rather than to the hadron.} as that in the Collins effect \cite{Collins:1992kk,Barone:2001sp,Barone:2005au}. The physical origin of the two asymmetries is, however, quite different. The Collins effect arises from final state interactions at the hadronization stage, which are incoherent with the interaction of the virtual photon. The rescattering giving rise to \eq{trres} is, on the other hand, coherent with the hard subprocess and thus can potentially affect also the total DIS cross section ($k_\perp$-integrated parton distributions). In particular, soft rescattering with a Pauli coupling eliminates an obstacle \cite{Barone:2001sp,Barone:2005au} to measuring transversity in DIS, namely that the upper part of the `handbag' conserves quark helicity.

The discussion of physical rescattering effects in \cite{Brodsky:2002ue} concerned the integrated parton distributions which are given by matrix elements of the form 
\beqa
f_{\qu/N}(x_B,Q^2) &=& \frac{1}{8\pi} \int dx^- \exp(-i x_B P^+ x^-/2) \nn
\\ &\times& \,
\left. \langle N(P)|
\bar\psi(x^-) \gamma^+\, W[x^-;0] \, \psi(0)|N(P)\rangle\right|_{x^+ = x_\perp =0}  
\label{PDFdef} 
\eeqa
where in the Bjorken limit all fields are evaluated at equal LF time 
$x^+ \sim 1/\nu$ and vanishing transverse separation $x_\perp\sim 1/Q$. In Feynman gauge the rescattering effects are given by the Wilson line 
\beq
W[x^-;0] = {\rm P}\exp\left[ig_s\int_0^{x^-} dw^- A^+(w^-) \right] \label{POE} 
\eeq
In light-cone gauge (lcg., $A^+=0$) the Wilson line reduces to unity,
\beq \label{POElcg}
W[x^-;0] = 1\ \ \ \mbox{ for } A^+=0
\eeq
which appears to suggest that there is no effect of rescattering in this gauge (and hence in any gauge, due to the gauge invariance of physical quantities).

The model analysis of \cite{Brodsky:2002ue} was done in both the Feynman and light-cone gauges. Diagrams involving rescattering of the struck quark were indeed found not to contribute in lcg., in accordance with \eq{POElcg}. Nevertheless, the same physical effects on the DIS cross section were found in both gauges. The apparent contradiction is resolved by the peculiar origin of the rescattering effect in light-cone gauge: The contributing Feynman diagrams involve instantaneous (in $x^+$) interactions {\it between spectators}, mediated by $A^+$ gluons which carry $k^+=0$. These contributions arise from the pole at $k^+=0$ of the gluon propagator in lcg.\footnote{The poles at $k^+=0$ are spurious: they cancel in the sum of all diagrams. However, the poles in the subset of diagrams which involve only spectator-spectator interactions give a non-vanishing contribution.},
\beq \label{lcprop}
d_{LF}^{\mu\nu}(k) = \frac{i}{k^2+\ieps}\left[-g^{\mu\nu}+\frac{n^\mu k^\nu+ k^\mu
n^\nu}{k^+}\right]
\eeq
where $n=(0,2,\bs{0}_\perp)$.
Such interactions are ``zero-modes'' \cite{Brodsky:1997de,Yamawaki:1998cy} of the $x^+=0$ nucleon state in the matrix element of \eq{PDFdef}. States defined at $x^+=0$ are ambiguous with respect to zero-modes since signals can be transmitted on a LF surface. The zero modes of the nucleon LF states thus differ in the Feynman and light-cone gauges.

If the possibility is admitted that soft rescattering can change the quark helicity (and we do not see how to exclude it) then the Wilson line \eq{POE} does not fully describe the rescattering effect. The Wilson line arises from the $-ig_s\gamma^\mu$ coupling in \eq{paulivert} which at high energy conserves helicity.

Our observations are qualitatively consistent with earlier suggestions (see, \eg, \cite{Ellis:1978pe,Nachtmann:1983uz}) that non-perturbative (\eg, instanton) effects may jeopardize QCD factorization. In particular, Nachtmann \cite{Nachtmann:1983uz} has emphasized that timing arguments allow hard QCD subprocesses to be influenced by soft interactions. In our example the soft exchanges carry finite momentum in the target rest frame. Consequently they form in a finite time, compatible with the finite coherence length \eq{ioffe} of the virtual photon.

Ellis \etal\ \cite{Ellis:1978pe} argued that parton distributions obey the usual DGLAP evolution in $Q^2$ even if they are not universal. This is consistent with the observation in \cite{Brodsky:2004hi} that the soft rescattering is independent of hard gluon emission. Partons produced at the hard vertex separate with a transverse velocity $v_\perp \sim Q/\nu$ which vanishes in the Bjorken limit. Soft rescattering occurring within the finite coherence length \eq{ioffe} therefore cannot resolve hard gluon emission and does not affect the evolution in $Q^2$.

\acknowledgments
We thank Stan Brodsky for helpful discussions. This research was supported in part by the
Academy of Finland through grant 102046. MJ acknowledges a grant from GRASPANP,  the Finnish Graduate School in Particle and Nuclear Physics.

\end{document}